\documentclass[aps,pra,twocolumn,longbibliography,footinbib]{revtex4-1}
\usepackage{graphicx}
\usepackage{amsmath}
\usepackage{amssymb}

\usepackage{color}
\usepackage{ulem}
\usepackage{bm}  

\newcommand {\scat} {\mathrm{sc}}
\newcommand  {\Ein}   {\mathrm{Ein}}
\newcommand  {\T}   {\mathrm{T}}

\begin{document}

\title{Comparison of three approaches to light scattering by dilute cold atomic ensembles}

\author{Igor M. Sokolov${}^{1,2}$}
\affiliation{$^{1}$Department of Theoretical Physics, Peter the Great St.-Petersburg Polytechnic University, 195251, St.-Petersburg, Russia \\
$^{2}$Institute for Analytical Instrumentation, Russian Academy of Sciences, 198103, St.-Petersburg, Russia}

\author{William Guerin}
\affiliation{Universit\'{e} C\^{o}te d'Azur, CNRS, Institut de Physique de Nice, France}

\begin{abstract}
Collective effects in atom-light interaction is of great importance for cold-atom-based quantum devices or fundamental studies on light transport in complex media. Here we discuss and compare three different approaches to light scattering by dilute cold atomic ensembles. The first approach is a coupled-dipole model, valid at low intensity, which includes cooperative effects, like superradiance, and other coherent properties. The second one is a random-walk model, which includes classical multiple scattering and neglects coherence effects. The third approach is a crude approximation only based on the attenuation of the excitation beam inside the medium, the so-called ``shadow effect''. We show that in the case of a low-density sample, the random walk approach is an excellent approximation for steady-state light scattering, and that the shadow effect surprisingly gives rather accurate results at least up to optical depths on the order of 15.
\end{abstract}

\maketitle

\section{Introduction}

Laser-cooled atomic samples are one of the main tools of fundamental research in atomic physics nowadays\,\cite{Fallani:2015,Ketterle:2015,Chang:2018}. They can serve, for instance, as toy models to develop or tests ideas and concepts for quantum information or to study condensed-matter phenomena in new regimes. In the recent years they have also become used in practical applications such as clocks, gravity sensors, etc.

For many of these applications, having more atoms in the sample allows increasing the signal-to-noise ratio. However large densities or large optical depths come along with potential new systematic effects. In particular, it has been recognized as early as 2004 that collective shifts due to the dipole-dipole interaction between atoms might become a limitation for atomic clocks\,\cite{Chang:2004}. More generally, many applications of cold atoms involve their interaction with optical fields and it is thus of fundamental importance to understand the collective effects in light-atom interactions.

There has been a lot of research on this topic in the recent years, especially in the weak-intensity regime (linear-optics or single-photon regime); see, for example, the reviews\,\cite{Guerin:2017a,Kupriyanov:2017}. Among the most important results, one can mention the difficulty of understanding the observed spectrum (shift and line shape) of light transmitted through dense clouds (see\,\cite{Jennewein:2018} and references therein) and the observation of super- and subradiance in the decay dynamics of light in dilute clouds\,\cite{Guerin:2016a,Araujo:2016,Roof:2016,Solano:2017,Weiss:2018}.

Several experiments have also been performed in a simpler situation, dealing with steady-state scattering from dilute and large clouds\,\cite{Labeyrie:2004,Bienaime:2010,Chabe:2014,Kemp:2018}, and those experiments have been interpreted in various ways, depending on which theoretical model was used for comparison.

In this article we would like to discuss and compare three possible approaches to describe this kind of experiments on steady-state ``incoherent'' scattering, i.e., off-axis scattering (out of the forward, coherent, transmitted beam). The first
model is the coupled-dipole (CD) model, which has been heavily used in the context of cooperative scattering\,\cite{Javanainen:1999,Svidzinsky:2010,Courteille:2010,Sokolov:2011,Kuznetsov:2011,Bienaime:2011,Bienaime:2013}. In this model the atomic dipoles form collective modes due to the dipole-dipole interaction mediated by light. Its strong advantage is that it is very complete, but this is also a drawback: it does not help much to understand the physics. We will thus compare this model with a ``random walk'' (RW) model, which considers photons bouncing from an atom to atom until leaving the sample. In this picture atoms are independent and only provide some scattering probability. We obtain that in the situation of linear optics, low density, steady-state off-axis scattering, this model is excellent. A preliminary comparison between these two models has been presented in\,\cite{Chabe:2014}. The main difference from work\,\cite{Chabe:2014} is that we take into account the vector nature of electromagnetic radiation and the Zeeman structure of atomic states with a $J=0 \rightarrow J=1$ transition. This allows us to conduct a more detailed comparison, including, in particular, the polarization properties of the scattered radiation, as well as its angular distribution. The good agreement between the results obtained in the CD and RW methods can serve as an additional justification for the latter, especially taking into account the polarization properties of light.

Finally we will also discuss a simple approach based on Beer-Lambert law, that is able to compute the total amount of scattered light (in all directions). Surprisingly, even if the scattered light is observed in one particular direction, this simple approach can provide quite accurate results up to moderate optical depth on the order of 15, with only a global free multiplicative factor for the intensity scale. This approach has already been used to explain recent experiments on the collective reduction of the radiation pressure force\,\cite{Chabe:2014,Bachelard:2016}, on off-axis scattering by very elongated clouds\,\cite{Kemp:2018}, and might also explain the effect reported in Ref.\,\cite{Machluf:2018}. In this work we analyze the applicability of this method in a wide range of parameters -- scattering angles, frequencies of the probe light, optical depths of the clouds.

\section{Description of the models}

\subsection{Coupled-dipole equations}

The coupled-dipole model has been widely used in the last years in the context of single-photon superradiance and subradiance~\cite{Svidzinsky:2010,Courteille:2010,Bienaime:2011,Bienaime:2013,Guerin:2016a,Roof:2016,Araujo:2016}.
It considers $N$ two-level atoms (ground state with the total angular momentum $J_g = 0$, and degenerate excited state $J_e = 1$ with $m\equiv J_z = -1,0,1$ ) at random positions $\bm{r}_i$ driven by an incident laser (Rabi frequency $\Omega(\bm{r})$, detuning $\Delta$).
Restricting the Hilbert space to the subspace spanned by the ground state of the atoms $|G\rangle = |g \cdots g \rangle$ and the singly-excited states $|i\rangle = |g \cdots e_{im} \cdots g\rangle$ and tracing over the photon degrees of freedom, one obtains an effective Hamiltonian describing the time evolution of the atomic wave function $| \psi(t) \rangle$,
\begin{equation}
| \psi(t) \rangle = \alpha(t) | G \rangle +  \sum\limits_{i=1}^N \sum\limits_{m} \beta_{e_{im}}(t)| i \rangle \; . \label{eq:psi}
\end{equation}
Considering the low intensity limit, when atoms are mainly in their ground states, i.e. $\alpha \simeq 1$, the problem amounts to determine the amplitudes $\beta_{e_{im}}$, which are then given by the linear system of coupled equations
\begin{equation}
\dot{\beta}_{e_{im}} = \left( i\Delta-\frac{\Gamma}{2} \right)\beta_{e_{im}} -\frac{i\Omega_{e_{im}}}{2} + \frac{i\Gamma}{2} \sum_{j \neq i}\sum\limits_{m'} V_{e_{im}e_{jm'}}\beta_{e_{jm'}}\; .
\label{eq:betas}
\end{equation}
These equations are the same as those describing $N$ classical dipoles driven by an oscillating electric field~\cite{Svidzinsky:2010}.
The first term on the right hand side corresponds to the natural evolution of independent dipoles (with $\Gamma$ the linewidth of the transition), the second one to the driving by the external laser, the last term corresponds to the dipole-dipole interaction and is responsible for all collective effects.  It reads
\begin{eqnarray}
V_{e_{im}e_{jm'}}& =&
-\frac{2}{\Gamma} \sum\limits_{\mu, \nu}
\mathbf{d}_{e_{im} g_i}^{\mu} \mathbf{d}_{g_j e_{m'}j}^{\nu}
\frac{e^{i k_0 r_{ij}}}{\hbar r_{ij}^3}
\nonumber
\\
&\times& \left\{
\vphantom{\frac{r_{ij}^{\mu} r_{ij}^{\nu}}{r_{ij}^2}}
 \delta_{\mu \nu}
\left[ 1 - i k_0 r_{ij} - (k_0 r_{ij})^2 \right]
\right.
\nonumber \\
&-&\left. \frac{\mathbf{r}_{ij}^{\mu} \mathbf{r}_{ij}^{\nu}}{r_{ij}^2}
\left[3 - 3 i k_0 r_{ij} - (k_0 r_{ij})^2 \right]
\right\}.
\label{eq:green}
\end{eqnarray}
Here $\mathbf{d}_{e_{im} g_i}$ is the dipole moment operator of the transition ${g} \to {e_m}$ of the atom $i$, $\mathbf{r}_{ij} =\mathbf{r}_i - \mathbf{r}_j$, $r_{ij} = |\bm{r}_i - \bm{r}_j|$ and
$k_0=\omega_0/c$ is the wavevector associated to the transition, with $c$ the vacuum speed of light. The superscripts $\mu$ or $\nu$ denote projections of vectors on one of the axes $\mu$, $\nu = x$, $y$, $z$ of the reference frame.

From the computed values of $\beta_{e_{im}}$, we can derive the intensity $I_\alpha (\boldsymbol{\Omega},t )$ of the light  polarization component $\alpha$ radiated by the cloud in a unit solid angle around an arbitrary direction given by  the wave vector $\mathbf{k}$ of scattered light ($\boldsymbol{\Omega}={\theta,\varphi}$). For the steady state case it can be determined as follows\,\cite{Kuraptsev:2017}:
\begin{equation}
I_\alpha(\boldsymbol{\Omega})=\frac{c}{4\pi} \left|  k_{0}^{2} \sum\limits_{i,m}\left(\mathbf{u}_\alpha^*\mathbf{d}_{g_ie_{im}}\right)  \beta_{e_{im}} \exp \left(
-i\mathbf{k}\mathbf{r}_{i}\right)  \right| ^{2}.
\end{equation}
Here $\mathbf{u}_\alpha$ is the unit polarization vector of the scattered light.

The advantage of the CD model is that it is very complete, as it includes diffraction/refraction/attenuation effects due to the complex index of refraction, multiple scattering including coherence effects such as coherent backscattering (CBS), and super/sub-radiance. However it is computationally limited to a few thousand atoms, although a technic has been proposed to overcome this limitation\,\cite{Sutherland:2016}. It can be readily applied to experiments involving a very small number of atoms, like the ones of Ref.\,\cite{Jennewein:2018}, otherwise the parameters have to be scaled in an appropriate way\,\cite{Sokolov:2013}, which depends on the physics. It is thus important to know if the studied phenomena depend on the atomic density, on the resonant optical thickness, or on something else~\cite{Guerin:2017a}. Another limitation is that it is rather difficult to extend this approach to multilevel systems, even if there has been recently some progress in that direction~\cite{Lee:2016,Hebenstreit:2017,Sutherland:2017b,Munro:2018}.

As the CD model is very complete, it does not always allow identifying the most relevant physical interpretation of the observations. It is thus useful to make comparisons with other models, in which some physical ingredients are neglected.

\subsection{Random walk model}

The RW model provides such a possibility. This model is based on a ``photon'' picture (although no quantized optical field is required) propagating inside the atomic cloud. Coherent effects such as diffraction or refraction inside the cloud are neglected, as well as super- and subradiance. The atoms only provide scattering, characterized by a local mean-free path, depending on the density distribution and the atomic cross-section. This model is appropriate to compute average ``incoherent'' scattering, where here incoherent means that the phase is randomized by the configuration averaging, which is true in all directions except the forward one (coherent transmission). The scattering can still be elastic and keep the phase information along the path, and one can even recover the coherent back scattering cone\,\cite{Labeyrie:2003b,Kupriyanov:2003}.

There are two possible methods for simulating this random walk model, with different advantages and drawbacks from a numerical point of view, but they contain the same physics.

The first one is to use stochastic simulations\,\cite{Molisch} in which the fate of a photon is followed until it escapes the medium. At each scattering a random direction is drawn as well as a step length depending on the atomic density distribution and scattering cross-section. Averaging over many photons is then performed. Such simulations
have been used in many previous work, for steady-state scattering\,\cite{Labeyrie:2004,Chabe:2014,Eloy:2018} and temporal dynamics (radiation trapping)\,\cite{Labeyrie:2003,Labeyrie:2005,Balik:2005,Weiss:2018}. This method allows one to take into account the frequency redistribution induced by the temperature and one does not need to truncate the number of scattering orders.

Another method of implementation is different types of diagram techniques for nonequilibrium systems, as described in Refs.\,\cite{Labeyrie:2003b,Datzyuk:2006} (for more details see also the review\,\cite{Kupriyanov:2017}). This approach, which we use in the following, allows us to obtain an explicit expression for the scattered light intensity $I_\alpha(\boldsymbol{\Omega} )$ in the form of a series over the number of incoherent scattering events:
\begin{equation}
\label{2}
I_\alpha(\boldsymbol{\Omega} )=\sum\limits_{s} I^{(s)}_\alpha(\boldsymbol{\Omega} ) \,.
\end{equation}
Each term of this series is the multiple integral over coordinates (and velocities for moving atoms) of the atoms forming the corresponding atomic chain. As example we show below the contribution of double incoherent scattering
\begin{gather}
      I^{(2)}_\alpha(\boldsymbol{\Omega} )=\frac{ck_0^8}{4\pi}
      \int \!\!\,\int \!d^{3}r_{1}d^{3}r_{2}\dfrac{\rho(\mathbf{r}_{1})\rho(\mathbf{r}_{2})}
      {\left | \mathbf{\mathbf{r}_{1}}-\mathbf{\mathbf{r}_{2}}\right |^{2}}\times
      \label{3}  \\
      \left|\sum\limits_{\beta,\gamma,\delta,\mu,\nu} X_{\alpha\beta}(\mathbf{r},\mathbf{r}_{2},\omega)
      \chi_{\beta\gamma}(\omega)
      X_{\gamma\delta}(\mathbf{r}_{2},\mathbf{r}_{1},\omega)\times
       \nonumber \right.\\
  \left.  \chi _{\delta\mu}(\omega)
      X_{\mu\nu}(\mathbf{r}_{1}\mathbf{,r}_{0}\mathbf{,}\omega )
      \mathbf{u}_{\nu }E_0 \right|^2  . \notag
\end{gather}%
This expression has a fairly clear physical meaning. The function $X_{\mu\nu}(\mathbf{r}_1,\mathbf{r}_0,\omega)$ describes the propagation of light from the source $\mathbf{r}_{0}$ to the point $\mathbf{r}_1$ where the first incoherent scattering event takes place. The explicit form of this function is determined by the processes of coherent forward scattering in an optically dense but isotropic medium \cite{Labeyrie:2003b,Datzyuk:2006,Kupriyanov:2017}
\begin{eqnarray}
\label{4} X_{\mu\nu}(\mathbf{r}_1,\mathbf{r}_0,\omega)=\delta_{\mu\nu}\exp\left(-\frac{i b_0(\mathbf{r}_1,\mathbf{r}_0)}{2}\frac{\Gamma/2}{\omega-\omega_0+i\Gamma/2}\right),
\end{eqnarray}
where the resonant optical depth of the inhomogeneous medium between any arbitrary points $\mathbf{r}_1$ and $\mathbf{r}_2$ is
\begin{eqnarray}
\label{5}  &&b_0(\mathbf{r}_2,\mathbf{r}_1)=\sigma_{0}\int_{\mathbf{r}_1}^{\mathbf{r}_2}\rho(\mathbf{r})d\mathbf{s}.
\end{eqnarray}
Here $\sigma_0=6\pi/k^2$ is the resonant cross-section for the considered case of a $J=0\leftrightarrow J=1$ transition and $\rho(\mathbf{r})$ is the local atomic density.

At point $\mathbf{r}_1$ the light  undergoes incoherent scattering. Its propagation direction changes and its polarization also can change. This process is depicted by the scattering amplitude
\begin{equation}
      \chi _{\delta\mu}(\omega)\;=\;-\sum_{m}\,\frac{(d_{\delta})_{ge_m}\,(d_{\mu})_{e_mg}}{\hbar (\omega -\omega _{0})\,+i\hbar \Gamma /2}\, .  \label{6}
\end{equation}%
Then the photon propagates toward point $\mathbf{r}_2$ [function $X_{\gamma\delta}(\mathbf{r}_{2},\mathbf{r}_{1},\omega)/\left | \mathbf{\mathbf{r}_{1}}-\mathbf{\mathbf{r}_{2}}\right |$] and after the second incoherent scattering [$\chi_{\beta\gamma}(\omega)$] is sent to photodetector [$X_{\alpha\beta}(\mathbf{r},\mathbf{r}_a,\omega)$], whose position $\mathbf{r}$ is determined by direction $\boldsymbol{\Omega} $. Integration over $\mathbf{r}_1$ and $\mathbf{r}_2$ with weight $\rho(\mathbf{r}_{1})\rho(\mathbf{r}_{2})$ takes into account all possible acts of double incoherent scattering.

The contributions of higher scattering orders will contain higher order integrals, whose integrands will additionally contain factors corresponding to scattering amplitudes $\chi_{ii'}(\omega)$ and photon propagators $X_{jj'}(...)$. The computation of multiple integrals for each scattering order is performed using statistical Monte-Carlo methods. The total number of items that should be taken into account in the sum (\ref{2}) depends on the optical depth of the atomic clouds and is determined every time in the calculation process. Usually this number is twice as big as the optical depth.

As a conclusion of this section we note that the contributions like (\ref{3}) correspond to the so-called ladder diagrams and does not take into account interference under multiple scattering in optically thick ensembles. This interference can be included by taking into account the contribution of cross diagrams\,\cite{Labeyrie:2003b,Datzyuk:2006,Kupriyanov:2017}. However in the case of dilute media, which we consider here, it influences the light intensity only in a narrow cone near the backward scattering direction (CBS effect).

These two methods for implementing the RW model can be shown to be strictly equivalent. The difference is only in the computational features of each method, namely in the way of simulating the random walk of photon in the medium. In the first method, the number of scattering events experienced by a given photon as well as its escape direction and its polarization are random quantities. With the second method we consider scattering of different orders separately and we can calculate scattering in a given polarization channel and in a given direction. Moreover we can use the so-called ``essential sample'' method as part of the Monte Carlo procedure. All these advantages of the second method can significantly reduce the computation time. For these reasons in the present work all calculations using the RW model are performed by means of the second method.

\subsection{Beer-Lambert law}

The two previous models are very versatile as they can be used in various situations, including temporal dynamics\,\cite{Weiss:2018} and fluctuations\,\cite{Eloy:2018}. However, as far as steady-state scattering is concerned, it is sometimes useful to make use of the obvious result that the total scattered light equals the amount of light removed from the driving beam. This attenuation can be computed from Beer-Lambert law,
\begin{equation}
\frac{I_\T(x,y)}{I_0(x,y)} = \exp\left[- \sigma_\scat \int \rho(x,y,z) dz\right] \, ,
\end{equation}
where $I_0$, $I_\T$ are the incident and transmitted intensities, the driving beam propagates along $z$, $\rho(x,y,z)$ is the atomic density distribution and $\sigma_\scat$ is the scattering cross-section. The argument of the exponential is called the optical depth. A usual density distribution is a Gaussian with rms radius $R$ along $z$, which gives an optical depth
\begin{equation}
\begin{split}
b(x,y) & = \sigma_\scat\int\limits_{-\infty}^{\infty}\rho(x,y,0)\exp(-z^2/2R^2) dz \\
 & = \sqrt{2\pi}\rho(x,y,0)\sigma_\scat R \, .
\end{split}
\end{equation}

If the initial beam profile also has a simple intensity distribution, one can integrate over the transverse directions to get the total transmission, and thus the total attenuation, which is also the total scattered light. For instance, supposing the incoming beam to be a plane wave and the atomic cloud to be a Gaussian sphere (rms size $R$ in all directions, peak density $\rho_0$, peak optical depth $b=\sqrt{2\pi}\rho_0\sigma_\scat R$), one obtains (see\,\cite{Kemp:2018,Bachelard:2016}) that the total scattering cross-section of the cloud is
\begin{equation}\label{eq.Sigma_RW}
\Sigma_\scat = N \sigma_\scat \times \frac{\Ein(b)}{b} \, ,
\end{equation}
where $N$ is the atom number and Ein is the integer function~\cite{Wolfram:Ein}
\begin{equation}
\begin{split}
\Ein(b) & = \int_0^b \frac{1-e^{-x}}{x} \, dx \\
& = b \left[ 1 + \sum_{n=1}^\infty \frac{(-b)^n}{(n+1)(n+1)!} \right] \,.
\end{split}
\end{equation}
The factor $\Ein(b)/b$ in Eq.\,(\ref{eq.Sigma_RW}) corresponds to the deviation from single-atom physics.
In the limit of vanishing optical depth $b$, the value expected from single atom physics is recovered, $\Sigma_\scat=N\sigma_\scat$. For high optical depth, the cross-section increases only logarithmically, which appears as a collective saturation of the scattered light. This saturation comes from the fact that the atoms in the back of the cloud are less illuminated due to the extinction of the light caused by the destructive interference between the incident and the scattered fields (``shadow effect")\,\cite{Chabe:2014,Guerin:2017a,Kemp:2018,Bachelard:2016}.

This result tells nothing about the angular distribution of the scattered light, but it is an excellent approximation for the total scattered light. It only neglects refraction and diffraction effects, which could have an impact for very small clouds\,\cite{Jennewein:2018}. It also neglects the forward coherent lobe~\cite{Scully:2006,Bromley:2016,Roof:2016}, which actually is diffracted/refracted light (this light is also responsible for the extinction paradox\,\cite{Bienaime:2014}). The Beer-Lambert result for the total intensity is equivalent to integrating in all directions the scattered intensity computed from the RW model.

Since the emission diagram of the cloud is not included in this approach, it might seem useless for measurements performed in a given direction. For instance, let's take a slab: as the optical depth increases and reach high values, the diffuse transmission tends to zero while the diffuse reflection goes to one~\cite{Labeyrie:2004}. Nevertheless, in many experiments the range of explored optical depth is not very large, and the geometry of the cloud not so drastic. For instance, with a Gaussian cloud illuminated by a plane wave, the optical depth goes smoothly to zero on the edges, and those edges usually contribute significantly to the experimental signal. As a consequence, the  variation of the emission diagram with the optical depth is not so pronounced, and the scaling provided by Eq.\,(\ref{eq.Sigma_RW}) can still be a good approximation, provided a reference point to set the intensity scale.

\section{Comparisons and discussion}

In this section we show several comparisons between the three approaches. We will first compare the coupled-dipole and the random walk models in the case of a simple atomic structure. We will see that the agreement is excellent, which gives us confidence to assume the applicability of RW methods for atoms with a more complex level structure, as also demonstrated in the context of CBS\,\cite{Labeyrie:2003b}. Then we will compare the RW to the Beer-Lambert results.

\subsection{Coupled dipoles and random walk}

\begin{figure*}
\centering
\includegraphics[width=1.6\columnwidth]{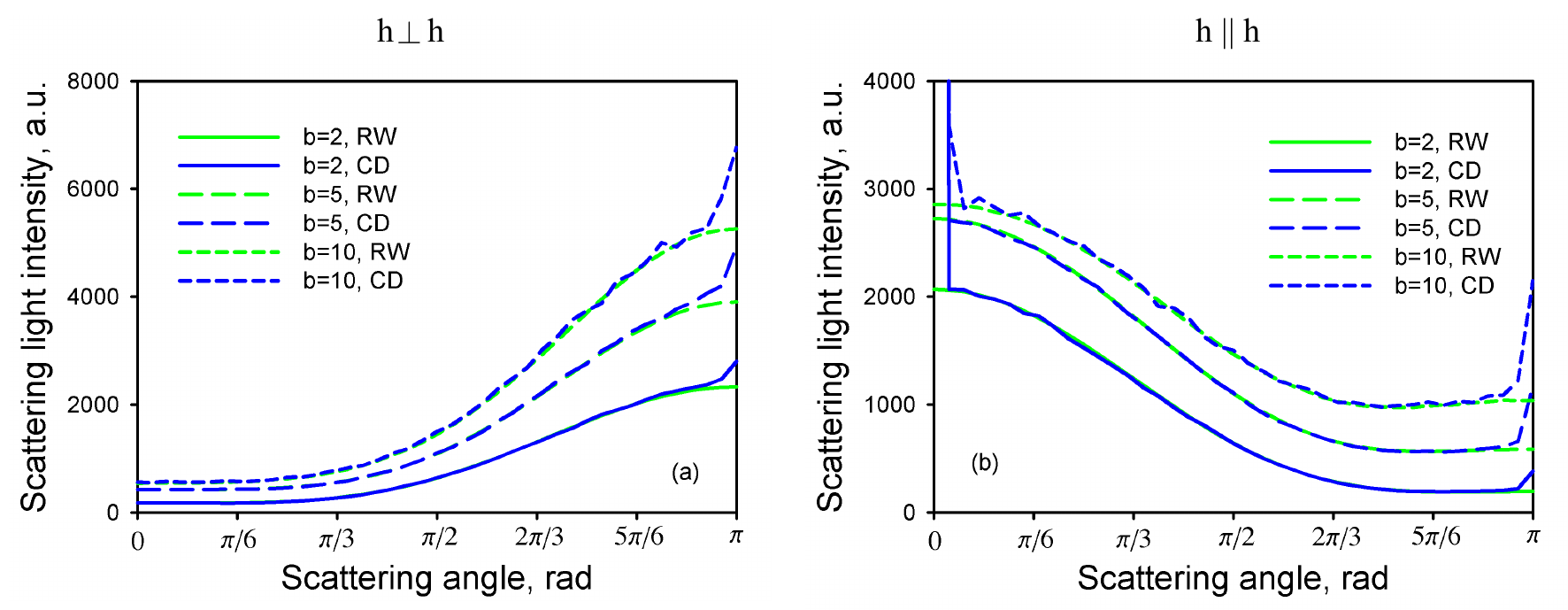}
\caption{Emission diagram computed with the coupled dipole (CD) and random walk (RW) models, for the orthogonal (a) and parallel (b) helicity channels, for several optical depths $b$. The laser is on resonance. The atomic density has a spherical Gaussian distribution of rms size $kR=50$. The agreement is excellent except in the forward scattering lobe and in the CBS cone. Averaging in CD calculation is performed over at least $10^4$ different atomic configurations for each curve. The total number of random walks is $2\times 10^6$.}
\label{fig.em_diag}
\end{figure*}

We start by showing in Fig.~\ref{fig.em_diag} the emission diagram (scattered intensity as a function of the angle) computed for different optical depths $b$ with the vectorial CD model and the RW simulations including the polarization (we suppose a $J=0 \rightarrow J=1$ atomic transition and take into account the corresponding emission diagram of each scatterer). The two panels show the results for the orthogonal and parallel helicity channels. Here the RW simulations do not include the crossed diagrams, as a consequence the coherent back scattering cone and the coherent forward lobe are not present in the RW results. Apart from those two specific directions, the agreement between the two models is excellent. The difference between the curves is comparable with the errors of the computational procedures caused by the use of Monte Carlo methods, as well as the limitation of the number of considered orders of multiple scattering in the RW model. We have also checked that the agreement is as good in other polarization channels. This is thus a generalization, in the vectorial case, of the comparison already made in\,\cite{Chabe:2014}.

\begin{figure}
\centering
\includegraphics[width=0.8\columnwidth]{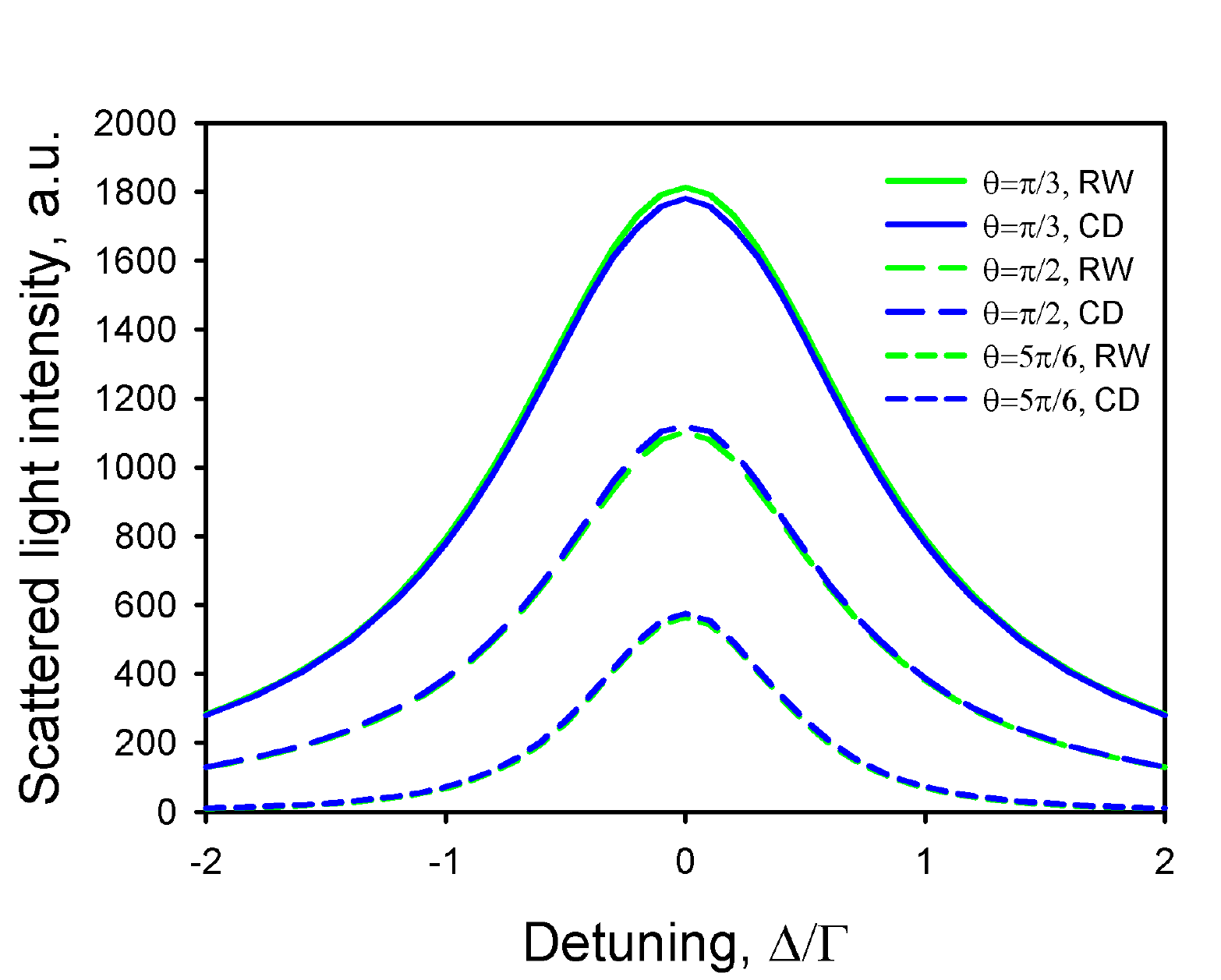}
\caption{Spectrum for several scattering angles, computed with the CD and RW models. The resonant optical depth is $b_0=5$ and the atomic sample is still Gaussian with $kR=50$. Averaging in CD calculation is performed over at least $5\times 10^4$ different atomic configurations for each curve. The total number of random walks is $2\times 10^6$.}
\label{fig.spectrum}
\end{figure}

Let us now turn to the spectrum. In Fig.~\ref{fig.spectrum} we compare the scattering spectrum for CD and RW simulations, with a given density and a given optical depth (dilute regime), for different scattering angles. Here also the agreement is excellent, the observed difference being compatible with the computational precision.

\begin{figure}
\centering
\includegraphics[width=0.8\columnwidth]{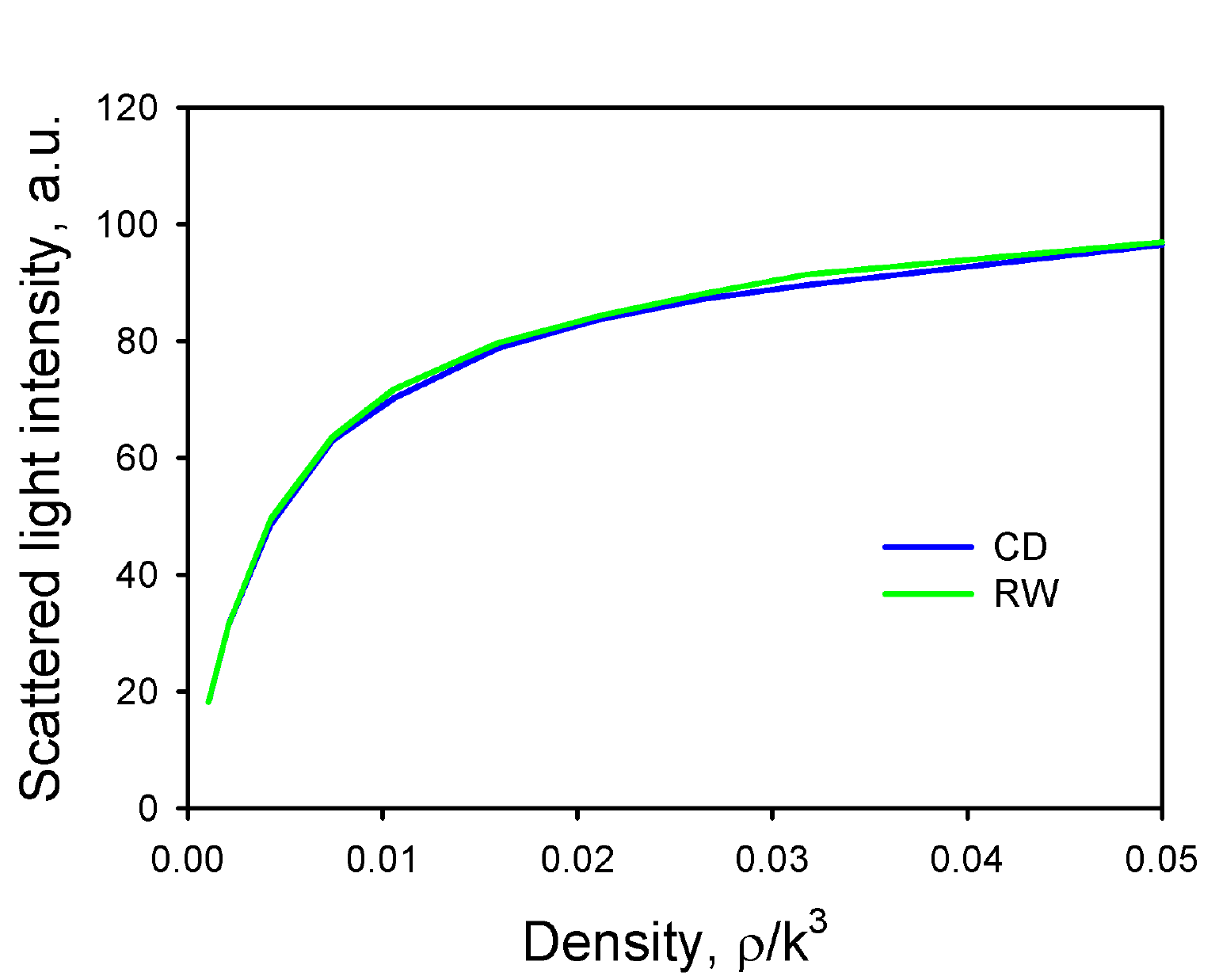}
\caption{Scattered intensity as a function of the density (the optical depth also varies). The scattering angle is $\theta=\pi/3$, the polarization channel is $h \parallel h$, the size of the sample is $kR=10$, the laser is at resonance. In this range of density (dilute cloud), no significant difference is visible between the CD and the RW models. The curves were averaged as in Fig.\,\ref{fig.em_diag}.}
\label{fig.density}
\end{figure}

In Fig.~\ref{fig.density} we explore the density dependence, but we keep a low density, i.e. $\rho k^{-3} \ll 1$. Here the size of the atomic sample is fixed ($kR=10$) so the density is varied by changing the atom number, which also changes the optical depth. Up to $\rho k^{-3} = 0.05$ we do not observe any significant difference between the two models.

From the three previous figures we conclude that at low density and for off-axis scattering, the two models give the same results. Note that in the absence of any temperature-induced frequency redistribution, the only relevant parameter in the RW model is the optical depth,
\begin{equation}
b(\Delta) = \frac{b_0}{1+4\Delta^2/\Gamma^2} \, ,
\end{equation}
where the resonant optical depth (for a Gaussian sample) is $b_0 = \sqrt{2\pi} \rho_0 \sigma_0 R = 3N/(kR)^2$ and the resonant scattering cross-section is $\sigma_0 = 3\lambda^2/(2\pi) = 6\pi/k^2$. As a consequence, in an experiment where several parameters can be varied (size, atom number, detuning), it is generally relevant to plot the experimental data as a function of the optical depth\,\cite{Kemp:2018}.

However we expect some discrepancies to appear at high densities, where refractive index effects and various collective shifts appear\,\cite{Jennewein:2018,Friedberg:1973,Scully:2009b,Manassah:2012,Roof:2015,Bromley:2016}. The precise study of these effects is beyond the scope of this paper.

\subsection{Random walk and Beer-Lambert}

\begin{figure*}
\centering
\includegraphics[width=1.6\columnwidth]{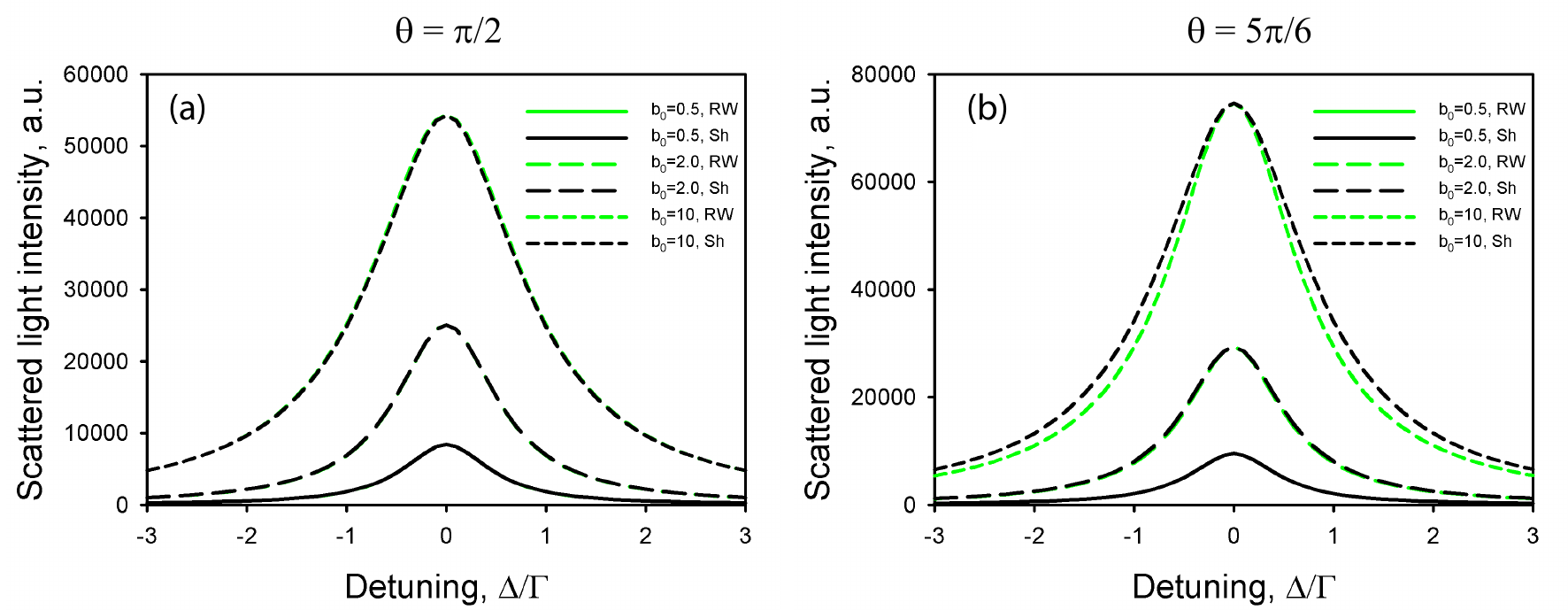}
\caption{Scattering spectrum at $\theta = \pi/2$ (a) and $\theta = 5\pi/6$ (b) computed with the RW and the shadow approximation, for several optical depths. The probe beam is circularly polarized and the cloud size is $kR=200$. What is plotted is the total intensity (summed over the polarizations). The results of the shadow model has been renormalized so that it coincides with RW calculations for $\Delta=0$. The number of different RWs is $2\times 10^6$. The agreement is perfect for $\theta = \pi/2$. The agreement is not as good for $b_0 = 10$ and $\theta=5\pi/6$ due to the anisotropy of the emission diagram at large optical depth due to multiple scattering. This effect, which increases the scattered intensity near the backward direction, is included in the RW model and not in the shadow effect.}
\label{fig.spectrum_shadow}
\end{figure*}

Since the RW model has been validated in the previous section, we now only compare the RW to the ``shadow effect'' (``Sh'' in figures) computed by Beer-Lambert law [Eq.\,(\ref{eq.Sigma_RW})]. We also make use of the possibilities to include a complicated level structure in the RW model: all results in this section were obtained for a $F=2 \rightarrow F=3$ transition with a statistical equipopulated mixture of the Zeeman ground states, typical of $^{87}$Rb experiments. Note that the two models yield results that only depend on the optical depth. However Eq.\,(\ref{eq.Sigma_RW}) only allows us to compute the total scattered light and not the radiation pattern. The purpose of this section is to see if it can still give useful results for scattered light detected in one particular direction.

For very low optical depths, multiple scattering is negligible and the emission diagram of the whole cloud is the same as the one of a single atom (e.g., isotropic in the scalar model), so the knowledge of the total cross-section is enough to compute the scattering in any direction.

For larger optical depths, the emission diagram is modified due to multiple scattering, with more light scattered in the backward directions and less light able to cross the sample~\cite{Labeyrie:2004}. However, for a Gaussian cloud, the difference appears slowly with the peak optical depth.

A first illustration is provided in Fig.\,\ref{fig.spectrum_shadow}, where two spectra are shown, one at $\theta = \pi/2$ (panel a) and one near the backward direction ($\theta = 5\pi/6$, panel b), for optical depths up to 10. The curves computed with the shadow effect have been vertically scaled to match the RW results on resonance. One can see that at $\theta = \pi/2$ the agreement is perfect. In the backward direction, there is only a slight discrepancy at large $b$. The difference in total width of the two curves for $b=10$ is about $10\%$, which is comparable with typical experimental uncertainties\,\cite{Kemp:2018}. Actually as the detuning is varied over the spectrum, the optical depth varies between its maximum on resonance and almost zero. The remarkable result is that the variation of the emission diagram is not big enough to significantly distort the spectrum line shape. Note that the two models would give identical results in the wings of the spectrum if one used a correct normalization, because there the optical depth is low. We also believe it would match experimental data without any free parameter if one could precisely calibrate the amount of detected light, accounting for the solid angle of the detection, the quantum efficiency of the detectors, etc.

\begin{figure}
\centering
\includegraphics[width=0.8\columnwidth]{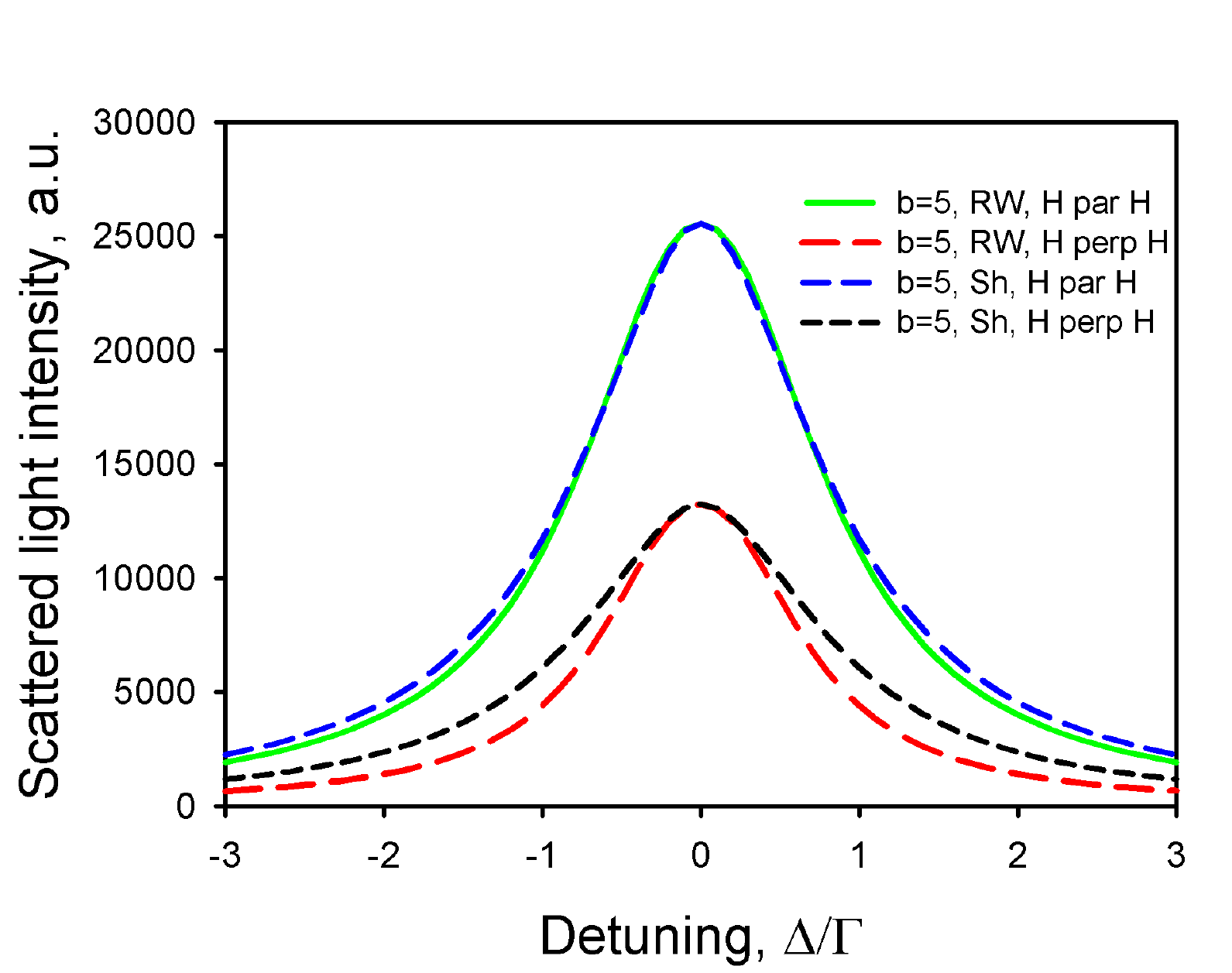}
\caption{Scattering spectrum computed for a $F=2 \rightarrow F'=3$ transition and for different helicity channels. The scattering angle is $\theta = \pi/3$ and the resonant optical depth is $b_0=5$. The number of different RWs is $2\times 10^6$. The results of the shadow model has been renormalized so that it coincides with RW calculations for $\Delta=0$.}
\label{fig.spectrum_shadow3}
\end{figure}

The figure\,\ref{fig.spectrum_shadow3} illustrates the same result, but we now address the polarization dependence. The scattering angle is $\theta = \pi/3$. With such an intermediate angle, a complex level scheme and the effect of multiple scattering near resonance (which randomizes the polarization), it is hard to have an intuitive prediction of the evolution of the emission diagram when the optical depth increases. The result of the comparison is that, still with a free multiplicative factor, the shadow effect gives very good results in the parallel helicity channel and fair results in the orthogonal one. Given the complexity of the RW model (with polarization and atomic levels), one can consider the Beer-Lambert method to be useful in obtaining quick and easy, but still accurate, results.

\begin{figure}
\centering
\includegraphics[width=0.8\columnwidth]{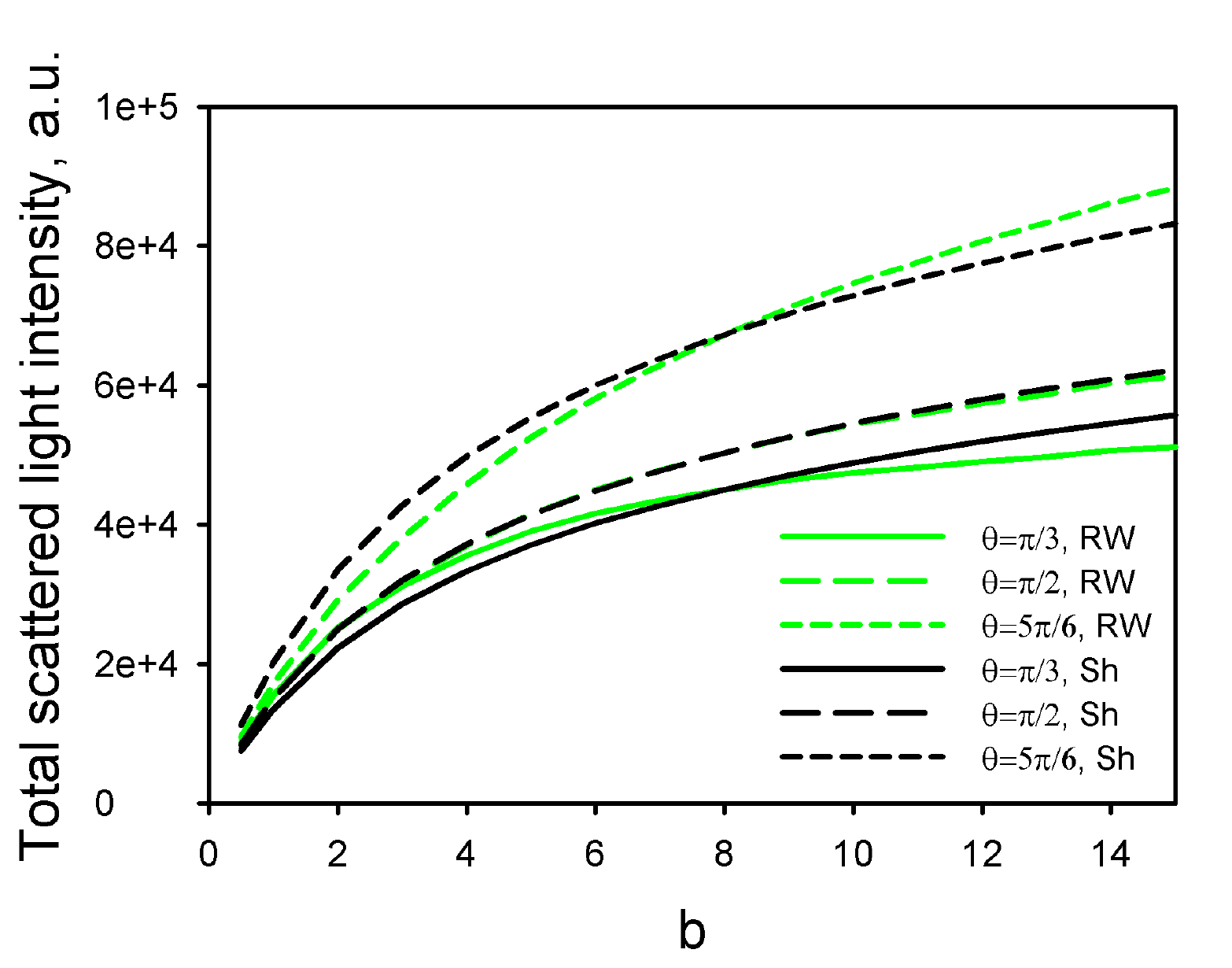}
\caption{Scattered light (summed over all polarizations) as a function of the optical depth for different scattering angles, computed with the RW and the shadow effect. The number of different RWs is $2\times 10^6$. The results of the shadow model has been renormalized so that it coincides with RW calculations for $b=7$.}
\label{fig.shadow_b}
\end{figure}

Finally, we plot in Fig.~\ref{fig.shadow_b} the scattered light at 3 different angles as a function of the optical depth $b$, from $b = 0.5$ to $b=15$. As previously, a free multiplicative factor is used to scale the data from the shadow effect. At $\theta = \pi/2$, the agreement with the RW model is excellent. At large $b$, fewer photons are scattered near the forward direction, so the shadow method overestimates the amount of scattering (curves at $\theta = \pi/3$), and the opposite happens near the backward direction (curves at $\theta = 5\pi/6$). Nevertheless, over the whole range of optical depth, the discrepancies are rather small.

\section{Conclusion}

To summarize, we have extended the work of\,\cite{Chabe:2014,Kemp:2018} by performing quantitative comparisons between the coupled-dipole model, a random walk approach that neglects coherent effects, and a very simple approximation, Beer-Lambert law, which allows computing the total scattered light. We considered only the simplest case of low density, large cloud, and off-axis scattering. We have obtained that in this situation, the RW model is in perfect agreement with the coupled-dipole model, and the Beer-Lambert law provides quite accurate results, at least up to resonant optical depths of 15, which is surprisingly good.

In this paper, we examined densities up to 0.05 in dimensionless units (i.e. in unit of $k^3$). For a wavelength of 780\,nm in absolute units, this corresponds to approximately $2.6\times10^{13}$\,cm$^{-3}$. This is high enough for mesoscopic ensembles of cold atoms. Thus, our calculation shows that the RW method is quite good in a wide range of densities.

For future works, it would be interesting to extend these comparisons, especially between the CD and RW models, at higher density. Obviously the two models should give different results because of finite size/diffraction/refraction effects, but also because the scattering properties may be changed due to near-field coupling between atoms and effects of recurrent scattering, whose role increases as the density of atoms increases. An adaptation of the RW model, following the works of Refs.\,\cite{Naraghi:2015,Naraghi:2016}, might be able to restore the agreement. The CD model could therefore serve as a benchmark for new approaches to multiple scattering of light in dense samples.

\section*{Acknowledgments}

The work was supported by the Russian Science Foundation (project 17-12-01085). We thank Mark Havey for useful discussions. WG thanks the cold-atom team of INPHYNI for many useful discussions. The results were obtained with the use of the computational resources of the supercomputer center at the Peter the Great St. Petersburg Polytechnic University.


\begin{thebibliography}{10}
\newcommand{\enquote}[1]{``#1''}

\bibitem{Fallani:2015}
L.~Fallani and A.~Kastberg, \enquote{Cold atoms: A field enabled by light,}
  Europhys. Lett. \textbf{110}, 53001 (2015).

\bibitem{Ketterle:2015}
W.~Ketterle, \enquote{Happy birthday {BEC},} Nature Phys. \textbf{11}, 982--983
  (2015).

\bibitem{Chang:2018}
D.~E. Chang, J.~S. Douglas, A.~Gonz\'ales-Tudela, C.-L. Hung, and H.~J. Kimble,
  \enquote{\textit{Colloquium}: Quantum matter built from nanoscopic lattices
  of atoms and photons,} Rev. Mod. Phys. \textbf{90}, 031002 (2018).

\bibitem{Chang:2004}
D.~E. Chang, J.~Ye, and M.~D. Lukin, \enquote{Controlling dipole-dipole
  frequency shifts in a lattice-based optical atomic clock,} Phys. Rev. A
  \textbf{69}, 023810 (2004).

\bibitem{Guerin:2017a}
W.~Guerin, M.~T. Rouabah, and R.~Kaiser, \enquote{Light interacting with atomic
  ensembles: collective, cooperative and mesoscopic effects,} J. Mod. Opt.
  \textbf{64}, 895--907 (2017).

\bibitem{Kupriyanov:2017}
D.~V. Kupriyanov, I.~M. Sokolov, and M.~D. Havey, \enquote{Mesoscopic coherence
  in light scattering from cold, optically dense and disordered atomic
  systems,} Phys. Rep. \textbf{671}, 1--60 (2017).

\bibitem{Jennewein:2018}
S.~Jennewein, L.~Brossard, Y.~R.~P. Sortais, A.~Browaeys, P.~Cheinet,
  J.~Robert, and P.~Pillet, \enquote{Coherent scattering of near-resonant light
  by a dense, microscopic cloud of cold two-level atoms: Experiment versus
  theory,} Phys. Rev. A \textbf{97}, 053816 (2018).

\bibitem{Guerin:2016a}
W.~Guerin, M.~O. Ara\'ujo, and R.~Kaiser, \enquote{Subradiance in a large cloud
  of cold atoms,} Phys. Rev. Lett. \textbf{116}, 083601 (2016).

\bibitem{Araujo:2016}
M.~O. Ara\'{u}jo, I.~Kre\v{s}i\'{c}, R.~Kaiser, and W.~Guerin,
  \enquote{Superradiance in a large cloud of cold atoms in the linear-optics
  regime,} Phys. Rev. Lett. \textbf{117}, 073002 (2016).

\bibitem{Roof:2016}
S.~J. Roof, K.~J. Kemp, M.~D. Havey, and I.~M. Sokolov, \enquote{Observation of
  single-photon superradiance and the cooperative {Lamb} shift in an extended
  sample of cold atoms,} Phys. Rev. Lett. \textbf{117}, 073003 (2016).

\bibitem{Solano:2017}
P.~Solano, P.~Barberis-Blostein, F.~K. Fatemi, L.~A. Orozco, and S.~L. Rolston,
  \enquote{Super-radiance reveals infinite-range dipole interactions through a
  nanofiber,} Nature Comm. \textbf{8}, 1857 (2017).

\bibitem{Weiss:2018}
P.~Weiss, M.~O. Ara\'ujo, R.~Kaiser, and W.~Guerin, \enquote{Subradiance and
  radiation trapping in cold atoms,} New. J. Phys. \textbf{20}, 063024 (2018).

\bibitem{Labeyrie:2004}
G.~Labeyrie, D.~Delande, C.~A. M\"uller, C.~Miniatura, and R.~Kaiser,
  \enquote{Multiple scattering of light in a resonant medium,} Opt. Commun.
  \textbf{243}, 157--164 (2004).

\bibitem{Bienaime:2010}
T.~Bienaim\'e, S.~Bux, E.~Lucioni, P.~W. Courteille, N.~Piovella, and
  R.~Kaiser, \enquote{Observation of cooperative radiation pressure in presence
  of disorder,} Phys. Rev. Lett. \textbf{104}, 183602 (2010).

\bibitem{Chabe:2014}
J.~Chab\'e, M.~T. Rouabah, L.~Bellando, T.~Bienaim\'e, N.~Piovella,
  R.~Bachelard, and R.~Kaiser, \enquote{Coherent and incoherent multiple
  scattering,} Phys. Rev. A \textbf{89}, 043833 (2014).

\bibitem{Kemp:2018}
K.~J. Kemp, S.~J. Roof, M.~D. Havey, I.~M. Sokolov, D.~V. Kupriyanov, and
  W.~Guerin, \enquote{Optical-depth scaling of light scattering from a dense
  and cold atomic $^{87}${Rb} gas,} arXiv:1807.10939  (2018).

\bibitem{Javanainen:1999}
J.~Javanainen, J.~Ruostekoski, B.~Vestergaard, and M.~R. Francis,
  \enquote{One-dimensional modeling of light propagation in dense and
  degenerate samples,} Phys. Rev. A \textbf{59}, 649 -- 666 (1999).

\bibitem{Svidzinsky:2010}
A.~A. Svidzinsky, J.~Chang, and M.~O. Scully, \enquote{Cooperative spontaneous
  emission of {$N$} atoms: Many-body eigenstates, the effect of virtual {Lamb}
  shift processes, and analogy with radiation of {$N$} classical oscillators,}
  Phys. Rev. A \textbf{81}, 053821 (2010).

\bibitem{Courteille:2010}
P.~W. Courteille, S.~Bux, E.~Lucioni, K.~Lauber, T.~Bienaim\'e, R.~Kaiser, and
  N.~Piovella, \enquote{Modification of radiation pressure due to cooperative
  scattering of light,} Eur. Phys. J. D. \textbf{58}, 69--73 (2010).

\bibitem{Sokolov:2011}
I.~Sokolov, D.~Kupriyanov, and M.~Havey, \enquote{Microscopic theory of
  scattering of weak electromagnetic radiation by a dense ensemble of ultracold
  atoms,} JETP \textbf{112}, 246 (2011).

\bibitem{Kuznetsov:2011}
D.~V. Kuznetsov, V.~K. Roerich, and M.~G. Gladush, \enquote{Local field and
  radiative relaxation rate in a dielectric medium,} JETP \textbf{113},
  647--658 (2011).

\bibitem{Bienaime:2011}
T.~Bienaim\'e, M.~Petruzzo, D.~Bigerni, N.~Piovella, and R.~Kaiser,
  \enquote{Atom and photon measurement in cooperative scattering by cold
  atoms,} J. Mod. Opt. \textbf{58}, 1942--1950 (2011).

\bibitem{Bienaime:2013}
T.~Bienaim\'e, R.~Bachelard, P.~W. Courteille, N.~Piovella, and R.~Kaiser,
  \enquote{Cooperativity in light scattering by cold atoms,} Fortschr. Phys.
  \textbf{61}, 377--392 (2013).

\bibitem{Bachelard:2016}
R.~Bachelard, N.~Piovella, W.~Guerin, and R.~Kaiser, \enquote{Collective
  effects in the radiation pressure force,} Phys. Rev. A \textbf{94}, 033836
  (2016).

\bibitem{Machluf:2018}
S.~Machluf, J.~B. Naber, M.~L. Soudijn, J.~Ruostekoski, and R.~J.~C. Spreeuw,
  \enquote{Collective suppression of optical hyperfine pumping in dense clouds
  of atoms in microtraps,} arXiv:1804.09759  (2018).

\bibitem{Kuraptsev:2017}
A.~S. Kuraptsev, I.~Sokolov, and M.~D. Havey, \enquote{Angular distribution of
  single photon superradiance in a dilute and cold atomic ensemble,} Phys. Rev.
  A \textbf{96}, 023830 (2017).

\bibitem{Sutherland:2016}
R.~T. Sutherland and F.~Robicheaux, \enquote{Coherent forward broadening in
  cold atom clouds,} Phys. Rev. A \textbf{93}, 023407 (2016).

\bibitem{Sokolov:2013}
I.~M. Sokolov, A.~S. Kuraptsev, D.~V. Kupriyanov, M.~D. Havey, and S.~Balik,
  \enquote{A scaling law for light scattering from dense and cold atomic
  ensembles,} J. Mod. Opt. \textbf{60}, 50--56 (2013).

\bibitem{Lee:2016}
M.~D. Lee, S.~D. Jenkins, and J.~Ruostekoski, \enquote{Stochastic methods for
  light propagation and recurrent scattering in saturated and nonsaturated
  atomic ensembles,} Phys. Rev. A \textbf{93}, 063803 (2016).

\bibitem{Hebenstreit:2017}
M.~Hebenstreit, B.~Kraus, L.~Ostermann, and H.~Ritsch, \enquote{Subradiance via
  entanglement in atoms with several independent decay channels,} Phys. Rev.
  Lett. \textbf{118}, 143602 (2017).

\bibitem{Sutherland:2017b}
R.~T. Sutherland and F.~Robicheaux, \enquote{Degenerated {Zeeman} ground states
  in the single-excitation regime,} Phys. Rev. A \textbf{96}, 053840 (2017).

\bibitem{Munro:2018}
E.~Munro, A.~Asenjo-Garcia, Y.~Lin, L.~C. Kwek, C.~A. Regal, and D.~E. Chang,
  \enquote{Population mixing due to dipole-dipole interactions in a
  one-dimensional array of multilevel atoms,} Phys. Rev. A \textbf{98} (2018).

\bibitem{Labeyrie:2003b}
G.~Labeyrie, D.~Delande, C.~A. M\"uller, C.~Miniatura, and R.~Kaiser,
  \enquote{Coherent backscattering of light by an inhomogeneous cloud of cold
  atoms,} Phys. Rev. A \textbf{67}, 033814 (2003).

\bibitem{Kupriyanov:2003}
D.~V. Kupriyanov, I.~M. Sokolov, P.~Kulatunga, C.~I. Sukenik, and M.~D. Havey,
  \enquote{Coherent backscattering of light in atomic systems: Application to
  weak localization in an ensemble of cold alkali-metal atoms,} Phys. Rev. A
  \textbf{67}, 013814 (2003).

\bibitem{Molisch}
A.~Molisch and B.~Oehry, \emph{Radiation Trapping in Atomic Vapours} (Oxford
  University Press, Oxford, 1998).

\bibitem{Eloy:2018}
A.~Eloy, Z.~Yao, R.~Bachelard, W.~Guerin, M.~Fouch\'e, and R.~Kaiser,
  \enquote{Diffusive wave spectroscopy of cold atoms in ballistic motion,}
  Phys. Rev. A \textbf{97}, 013810 (2018).

\bibitem{Labeyrie:2003}
G.~Labeyrie, E.~Vaujour, C.~A. M\"uller, D.~Delande, C.~Miniatura,
  D.~Wilkowski, and R.~Kaiser, \enquote{Slow diffusion of light in a cold
  atomic cloud,} Phys. Rev. Lett. \textbf{91}, 223904 (2003).

\bibitem{Labeyrie:2005}
G.~Labeyrie, R.~Kaiser, and D.~Delande, \enquote{Radiation trapping in a cold
  atomic gas,} Appl. Phys. B \textbf{81}, 1001--1008 (2005).

\bibitem{Balik:2005}
S.~Balik, R.~G. Olave, D.~I. Sukenik, M.~D. Havey, V.~M. Datsyuk, I.~M.
  Sokolov, and D.~V. Kupriyanov, \enquote{Alignment dynamics of slow light
  diffusion in ultracold atomic $^{85}$Rb,} Phys. Rev. A \textbf{72}, 051402(R)
  (2005).

\bibitem{Datzyuk:2006}
V.~M. Datzyuk and I.~M. Sokolov, \enquote{Coherent backscattering under
  conditions of pulsed radiation trapping,} JETP \textbf{102}, 724--736 (2006).

\bibitem{Wolfram:Ein}
E.~W. Weisstein, \enquote{Ein function,} From MathWorld--A Wolfram Web
  Resource. http://mathworld.wolfram.com/EinFunction.html.

\bibitem{Scully:2006}
M.~O. Scully, E.~S. Fry., C.~H. {Raymond~Ooi}, and K.~W\'odkiewicz,
  \enquote{Directed spontaneous emission from an extended ensemble of {$N$}
  atoms: Timing is everything,} Phys. Rev. Lett. \textbf{96}, 010501 (2006).

\bibitem{Bromley:2016}
S.~L. Bromley, B.~Zhu, M.~Bishof, X.~Zhang, T.~Bothwell, J.~Schachenmayer,
  T.~L. Nicholson, R.~Kaiser, S.~F. Yelin, M.~D. Lukin, A.~M. Rey, and J.~Ye,
  \enquote{Collective atomic scattering and motional effects in a dense
  coherent medium,} Nat. Commun. \textbf{7}, 11039 (2016).

\bibitem{Bienaime:2014}
T.~Bienaim\'e, R.~Bachelard, J.~Chab\'e, M.~Rouabah, L.~Bellando, P.~W.
  Courteille, N.~Piovella, and R.~Kaiser, \enquote{Interplay between radiation
  pressure force and scattered light intensity in the cooperative scattering by
  cold atoms,} J. Mod. Opt. \textbf{61}, 18 (2014).

\bibitem{Friedberg:1973}
R.~Friedberg, S.~R. Hartmann, and J.~T. Manassah, \enquote{Frequency shifts in
  emission and absorption by resonant systems of two-level atoms,} Phys. Rep.
  \textbf{7}, 101--179 (1973).

\bibitem{Scully:2009b}
M.~O. Scully, \enquote{Collective {Lamb} shift in single photon {Dicke}
  superradiance,} Phys. Rev. Lett. \textbf{102}, 143601 (2009).

\bibitem{Manassah:2012}
J.~T. Manassah, \enquote{Cooperative radiation from atoms in different
  geometries: decay rate and frequency shift,} Adv. Opt. Photon. \textbf{4},
  108--156 (2012).

\bibitem{Roof:2015}
S.~Roof, K.~Kemp, M.~D. Havey, I.~M. Sokolov, and D.~V. Kupriyanov,
  \enquote{Microscopic lensing by a dense, cold atomic sample,} Opt. Lett.
  \textbf{40}, 1137 (2015).

\bibitem{Naraghi:2015}
R.~{Rezvani~Naraghi}, S.~Sukhov, J.~J. S\'aenz, and A.~Dogariu,
  \enquote{Near-field effects in mesoscopic light transport,} Phys. Rev. Lett.
  \textbf{115}, 203903 (2015).

\bibitem{Naraghi:2016}
R.~{Rezvani~Naraghi} and A.~Dogariu, \enquote{Phase transition in diffusion of
  light,} Phys. Rev. Lett. \textbf{117}, 263901 (2016).

\end{thebibliography}

\end{document}